\documentclass[prb,amssymb,floatfix]{revtex4}
\usepackage[latin9]{inputenc}
\usepackage{graphicx,bm,}
\usepackage{color}
\usepackage{amssymb,amsfonts,amsmath}
\usepackage{graphicx,psfrag,xspace}
\usepackage{color}
\usepackage{natbib}
\usepackage{dcolumn}
\usepackage{bm}
\usepackage{subfigure}
\usepackage{setspace}
\usepackage{url}
\usepackage{xcolor}
\usepackage{version}
\usepackage{bbold} 
\usepackage{appendix}
\usepackage{epstopdf}
\usepackage{relsize}

\newcommand{\vecrho}{\boldsymbol{\rho}}
\newcommand{\veceta}{\boldsymbol{\eta}}

\newcommand{\grad}{\bm{\nabla}} 

\newcommand{\be}{\begin{equation}}
\newcommand{\ee}{\end{equation}}

\begin{document}

\title{Efficient sampling of knotting-unknotting pathways for semiflexible Gaussian chains}

\author{Cristian Micheletti$^1$ and Henri Orland$^{2,3}$}
\affiliation{$^1$ SISSA, via Bonomea 265, I-34136 Trieste, Italy\\
$^{2}$ Institut de Physique Th\'eorique, CEA, CNRS, UMR3681,
F-91191 Gif-sur-Yvette, France \\
$^3$ Beijing Computational Science Research Center, No.10 East Xibeiwang Road, Beijing 100193, China \\}

\begin{abstract}
We propose a stochastic method to generate exactly the overdamped Langevin dynamics of semi-flexible Gaussian chains, conditioned to evolve between given initial and final conformations in a preassigned time. The initial and final conformations have no restrictions, and hence can be in any knotted state. Our method allows the generation of statistically independent paths in a computationally efficient manner. We show that these conditioned paths can be exactly generated by a set of local stochastic differential equations. The method is used to analyze the transition routes between various knots in crossable filamentous structures, thus mimicking topological reconnections occurring in soft matter systems or those introduced in DNA by topoisomerase enzymes. We find that the average number of crossings, writhe and unknotting number are not necessarily monotonic in time and that more complex topologies than the initial and final ones can be visited along the route.
\end{abstract}
\date{27/05/2017}
\maketitle



\section{Introduction}

Filamentous systems are typically strongly affected by topological constraints in their conformational, mechanical and dynamical properties. This is especially evident for self-avoiding ring polymers, which are trapped in a specific knotted state that cannot be altered in the course of their free dynamical evolution.
For these reasons, much interest has been spurred recently by the dramatic topological changes observed in crossable filamentous structures that can appear in dissipative systems. Notable instances include entangled optical beams \cite{Desyatnikov2012,Kedia2013}, vortex lines in fluids \cite{Kleckner:2013gu}, magnetic field lines in a plasma \cite{Kedia2016} and defect lines in liquid crystals \cite{Tkalec2011,jampani2011colloidal,martinez2014,machon2013knots,irvine2014,campbell2014topological}.  By contrast to polymers, strand crossings can occur when these filaments collide (subject to specific local conservation rules \cite{Laing2015,Liu2016,Scheeler2014}), thus creating the conditions for dynamical changes in topology.

Significant efforts are being made to map out the possible reconnection pathways and establish their recurrence across various dissipative systems \cite{Taylor2016,Kleckner:2016jx}. A key related question is whether the observed modes of topological changes are any different from those sustained by semi-flexible phantom rings. These systems, in fact, serve as terms of reference to understand the action of topoisomerase enzymes. These are enzymes that can progressively simplify the global knotted topology of DNA rings by fostering suitable local strand passages. Many efforts are accordingly made to understand which local selection criteria for strand passage would have the same disentangling effects on knotted phantom rings \cite{SHISHIDO1987215,Ullsperger1996,Rybenkov1997,Hua2007,Grainge2007,Vologodskii2016}.

Advancements along these lines depend, at least in part, on the possibility to generate computationally, or predict theoretically, physically-viable trajectories connecting two conformations with preassigned topology. This task is, in general, very challenging because spontaneous dynamical evolutions from a given initial state are unlikely to end up in a preassigned target one within a finite computational time, especially when significant free energy barriers are present along the route. Such difficulties are usually tackled by accelerating the dynamics using path sampling methods~\cite{olender1996calculation,elber2013reaction,bello2015exact,pan2008finding,meng2016transition,banisch2015reactive,PhysRevLett.97.108101,Bolhuis2002,0295-5075-113-1-18004} or with steered molecular dynamics techniques based on suitable external, and~possibly time-dependent forces \cite{schlitter1993targeted,Grubmuller1996,Isralewitz2001,PhysRevE.93.052144,paci1999forced,camilloni2011hierarchy}.  These schemes have proved essential for profiling free energy landscapes and establishing the salient steps along transition pathways. At the same time, they usually do not leave good control over the probabilistic weight of the trajectories, and hence on their representative significance.

Here we present a novel theoretical, and computationally efficient scheme based on Langevin bridges \cite{orland2011generating,majumdar2015effective} that allows one to connect two states with preassigned geometry by means of unbiased and physically-viable trajectories. With this strategy, that is entirely general,  we are able to study in great detail the canonically-relevant transitions pathways of semi-flexible rings between two assigned conformations of any topology. We show that these canonical transition pathways are often not minimal, meaning that more complex topologies than the initial and final ones can be visited along the route. This exposes an unsuspectedly rich phenomenology of topological rearrangements that could be explored and verified in future experiments on entangled soft matter systems.

\section{Methods}
\unskip
\subsection{The Conditioned Langevin Equation}

For the sake of simplicity, we start by illustrating the method on a one-dimensional
system, following closely the presentation given in ref. \cite{orland2011generating}.
We assume that the system is driven by a force $F(x,t)$ and is subject to stochastic dynamics in the form of an overdamped
Langevin equation:

\begin{equation}
\frac{dx}{dt}=\frac{1}{\gamma}F(x(t),t)+ \eta(t)\label{eq:langevin}
\end{equation}
where $x(t)$ is the position of the particle at time $t$ which experiences
the force $F(x,t)$. The friction coefficient $\gamma$ is related
to the particle diffusion coefficient $D$ through the Einstein relation $D=k_{B}T/\gamma$,
where $k_{B}$ is the Boltzmann constant and $T$ the temperature
of the thermostat. Finally, $\eta(t)$ is a Gaussian white noise
with moments given by $\langle\eta(t)\rangle=0$ and $\langle\eta(t)\eta(t')\rangle= 2\, D\, \delta(t-t')$.

One can show (see refs.~\cite{orland2011generating}, and supplemental material) that the Langevin trajectories starting at $x=0$ at time $t=0$ and conditioned  to end at $x_f$ at time $t_f$, can
be generated by a Langevin equation with an additional potential
force
\begin{equation}
\frac{dx}{dt}=\frac{1}{\gamma}F+2D\frac{\partial\ln Q}{\partial x}+\eta(t)
\label{eq:bridge1}
\end{equation}
where
\begin{equation}
Q(x,t)= P(x_f,t_f|x,t)
\end{equation}
and $P(x_f,t_f|x,t)$ denotes the probability to find the particle at $x_f$ at time $t_f$, given that it was at $x$ at~time $t$.

This equation generates Brownian paths, starting at $(x_{0},0)$ conditioned to end at $(x_{f},t_{f})$, with
unbiased statistics. It is the additional term $2D\frac{\partial\ln Q}{\partial x}$
in the conditioned Langevin equation that guarantees that the trajectories starting
at $(x_0,0)$ will end at $(x_{f},t_{f})$  and are statistically unbiased.

Equation (\ref{eq:bridge1}) is straightforwardly generalized to systems with many degrees of freedom. Specifically,
for systems comprising $N$ particles interacting via a potential $U$ and subject to an
external force ${\bf F}_n$ acting on particle $n$, the evolution of the position vector ${\bf r}_n$ of the $n$th particle, is given by:
\begin{equation}
\frac{d {\bf r}_{n}}{dt}=-\frac{1}{\gamma} \grad_{{\bf r}_n} U+\frac{1}{\gamma}{\bf F}_n(t)+2D\grad_{{\bf r}_n} \ln Q(\left\{ {\bf r}_{n}\right\} ,t) +\veceta_{n}(t)\label{eq:bridge2-1}
\end{equation}
where $Q(\left\{ {\bf r}_{n}\right\} ,t)=P\left(\{ {\bf r}_{n}^{(f)}\} ,t_{f}\mid \{ {\bf r}_{n}\},t\right)$
and the Gaussian noise $\veceta_{n}(t)$ satisfies
\vspace{12pt}\begin{eqnarray}
&\langle\eta_{n}^{(\alpha)}(t)\rangle=0\label{eq:noise1-1}, \ \ \
&\langle\eta_{n}^{(\alpha)}(t)\eta_{n'}^{(\alpha')}(t')\rangle=\frac{2\, {k_{B} T}}{\gamma}\delta_{nn'}\delta_{\alpha\alpha'}\delta(t-t')\label{eq:noise2-1}
\end{eqnarray}
where $\alpha$ labels the Cartesian coordinates $x,y,z$.

\subsection{Polymer Chain}
We now specialize Equation~(\ref{eq:bridge2-1}) to the case of ring polymers that freely evolve under the action of the following inter-monomer potential, $U$
\begin{equation}
\frac{U}{k_B T}=\sum_{n=1}^{N} \left[ \frac{3}{2a^{2}}({\bf r}_{n+1}-{\bf r}_{n})^{2} \nonumber +\frac{K}{2}({\bf r}_{n+1}-2{\bf r}_{n}+{\bf r}_{n-1})^{2}\right] \nonumber
\end{equation}
where ${\bf r}_{N}={\bf r}_{0}$ and ${\bf r}_{N+1}={\bf r}_{1}$, since the chain is a ring.

The first term is the elasticity of the polymer chain, whereas the second represents its bending rigidity. This expression for the bending rigidity is approximate, since the monomer length is not fixed in this model. However, this is a standard mean-field type model to represent semi-flexible polymers. We further assume no external force, $F_n(t)=0$.

To model chains with preassigned root-mean-square bond length, $b$, and  persistence length, $l_{P}$, the bare parameters $a$ and $K$ must be set by solving the following equations:
\begin{equation}
\begin{array}{ll}
l_{P}=\sqrt{\frac{K a^2}{3}} \vspace{3pt}\\
b^{2}=\frac{1}{N}\langle\sum_{n=1}^{N}\left({\bf r}_{n+1}-{\bf r}_{n}\right)^{2}\rangle \vspace{3pt}\\
~~~~= \frac{a^2 }{N}\sum_{p=0}^{N-1} \left[1+\frac{2Ka^2}{3}(1-\cos\omega_{p}) \right]^{-1}
\label{eq:bond}
\end{array}
\end{equation}
where $\omega_{p}=\frac{2\pi}{N}p$. For large $K$ and sufficiently long chains, Equation~(\ref{eq:bond}) yields the expected linear dependence of the persistence length on the chain bending rigidity (see Supplemental Material).

For the considered polymer case, the Langevin bridge equation of Equation~(\ref{eq:bridge2-1}) is best expressed in Fourier space:
\begin{equation}
\frac{d \tilde{\vecrho}_p}{dt} = -\Omega_p \tilde{\vecrho}_p + \frac{D}{N} \grad_{\tilde{\vecrho}_p}  \log Q + \tilde{\veceta}_p
\end{equation}
where
\begin{eqnarray}
\tilde{\vecrho}_{p}&=&\frac{2}{N}\sum_{n=1}^{N}\cos(\omega_{p}n)\,{\bf r}_{n}  \label{fourier_inv}\\
\Omega_p &=& (3/a^2)\, (1-\cos \omega_p) + 2K (1-\cos \omega_p)^2
\end{eqnarray}
and  $\tilde \veceta_p$ are the Fourier series of $\veceta_n(t)$ and are thus Gaussian white noises, defined by their moments
\begin{eqnarray}
\langle \tilde \eta_p^\alpha(t) \rangle &=&0 \\
\langle \tilde \eta_0^{(\alpha)}(t) \tilde \eta_p^{(\alpha')}(t')\rangle &=&\frac{2D}{N} \delta_{p0} \delta_{\alpha \alpha'}\delta(t-t') \\
\langle \tilde \eta_p^{(\alpha)} (t) \tilde \eta_{p'}^{(\alpha')}(t') \rangle &=&\frac{D}{N} \delta_{pp'} \delta_{\alpha \alpha'} \delta(t-t')\ .
\end{eqnarray}

The Green's function $Q(\tilde{\rho}_{p},t)$ can be computed exactly
by solving the Langevin equation in Fourier space. The calculation is straightforward, see Supplemental material, and yields the following bridge equation

\begin{equation}
\label{bridge_pol1}
\frac{d \tilde \vecrho_0}{dt} =\frac{\tilde \vecrho_0^{(f)}-\tilde \vecrho_0(t)}{t_f-t}+\tilde \veceta_0(t)
\end{equation}
\begin{equation}
\label{bridge_pol2}
\frac{d \tilde \vecrho_p}{dt} =-\Omega_p \tilde \vecrho_p(t) \vspace{3pt}+ \frac{\Omega_p}{\sinh [\Omega_p (t_f-t)]}
\left( \tilde \vecrho_p^{(f)}- \tilde \vecrho_p(t)e^{-\Omega_p (t_f-t)}  \right) \vspace{3pt} +\tilde \veceta_p(t)
\end{equation}
where $\Omega_p = (3/a^2)\, (1-\cos \omega_p) + 2K (1-\cos \omega_p)^2$ and $\tilde \vecrho_p^{(f)}$ denotes the final configuration of the chain in Fourier components. These equations can be discretized and solved numerically, from an initial configuration $\tilde \vecrho_p^{(0)}$ to a final one $\tilde \vecrho_p^{(f)}$.

We point out that the time-reversed trajectory is a legitimate solution of the bridge equations starting from $\{ {\bf r}^{(f)}\}$ at time $t=0$ and ending in $\{ {\bf r}^{(0)}\}$ at time $t=t_f$.  Also note that within this model, the contour length of the chain is not conserved during the time evolution. For representation purposes, it is possible to rescale the contour length to its initial value at any given time $t$ when inverting back from Fourier to real space representation.

\subsection{Circular Permutations}

In the bridge Equations (\ref{bridge_pol1}) and (\ref{bridge_pol2}),  monomers in the initial and final states are in one-to-one correspondence. To study the evolution between two ring shapes in the absence of external forces, one should allow the initial configuration $\{{\bf r}_1^{(0)}, \ldots, {\bf r}_n^{(0)} \}$ to go to any circular permutation of the final configuration, i.e. $\{{\bf r}_{1+n_0}^{(f)}, \ldots, {\bf r}_{N+n_0}^{(f)} \}$, for any $n_0=0,\ldots,N-1$, where we assume periodic conditions since the chain is a ring ${\bf r}_{n+N}={\bf r}_n$.
This requires substituting the single final state with a combination of its circular permutations.
 It has been shown \cite{orland2011generating} that if the final state is a combination of several states, the function $Q$ should be modified as
\begin{equation}
Q(\left\{ r_{n}\right\} ,t)=\sum_{n_0=0}^{N-1} P\left(\{ r_{n+n_0}^{(f)}\} ,t_{f}\mid \{r_{n}\},t\right)
\end{equation}
\noindent so that in absence of an external force, the bridge equations become
\begin{equation}
\frac{d \tilde \rho_0}{dt} = \frac{\tilde \rho_0^{(f)}-\tilde \rho_0(t)}{t_f-t}+\tilde \eta_0(t)
\end{equation}
\begin{equation}
\frac{d \tilde \rho_p}{dt} =-\Omega_p \tilde \rho_p(t)+ \frac{\Omega_p}{\sinh [\Omega_p (t_f-t)]}
\frac{1}{Q} \sum_{n_0=1}^{N-1} \left( \tilde \rho_p^{(n_0)}-  \tilde \rho_p(t) e^{-\Omega_p (t_f-t)} \right)P_1^{(n_0)} + \tilde \eta_p(t) \ . \label{bridge_circ}
\end{equation}
where
\begin{equation}
\tilde \rho_p^{(n_0)} = \frac{2}{N}\sum_{n=1}^N \cos (\omega_p n)\ r_{n+n_0}^{(f)}
\end{equation}
and
\begin{equation}
P_1^{(n_0)}=\exp \Bigg(-\frac{N}{D}\sum_{p=1}^{N-1}\Omega_p\frac{ \left(\tilde \vecrho_p^{(n_0)} - \tilde \vecrho_p(t) e^{-\Omega_p (t_f-t)} \right)^2}{1-e^{-2 \Omega_p (t_f-t)} } \nonumber
\Bigg)
\end{equation}

Similarly to the case without circular permutation, these equations are easily solved by discretization. The numerical complexity is increased due to the summation over circular permutations in Equation (\ref{bridge_circ}).

\section{Results and Discussion}

We used the Langevin bridging scheme to connect various pairs of ring polymer conformations tied in different knot types. The initial and final structures were picked from an equilibrated distribution (generated with a Monte Carlo scheme) of self-avoiding semi-flexible rings. These were modelled as a succession of $N=240$ cylinders with diameter $\sigma=b/4$, where $b$ is the length of the cylinder axis, and nominal Kuhn length equal to $10b$. For integrating the dynamics, and presenting the results, we took $b$ as the unit of length, and $D^2/b$ as the unit of time. In these units, the dynamics was integrated with a time step equal to  $10^{-4}$ and for a total timespan equal to 2.

The excluded volume interactions between the cylinders were then switched off during the Langevin bridging dynamics to allow for topology-unrestricted interconversions. By doing so we model the interconversions observed for defect lines in liquid crystals or vortex lines in fluids in the simplest possible manner. In the mentioned systems, in fact, self-crossings events have an energy cost or are subject to local conservation laws. In this first study, we neglect such interactions to keep the model amenable to extensive theoretical treatment and hence clarify the physically-viable reconnections routes in the simplest and most general setup.

We first discuss the transition from an unknotted conformation to a knotted one, and specifically to a left-handed $5_1$ knot. This topology belongs to the family of torus knots, which are drawable without self-intersections on the surface of a torus \cite{Adams:2004uu}. We chose it as a first example, because it is the simplest knot type with unknotting number equal to 2. This means that, even in the most favorable conditions, the transition from the trivial to the $5_1$ topology cannot occur via a single strand passage, but at least two are needed. This ought to yield interesting knotting pathways.

\begin{figure}[h!]
\centering
\includegraphics[width=0.7\columnwidth]{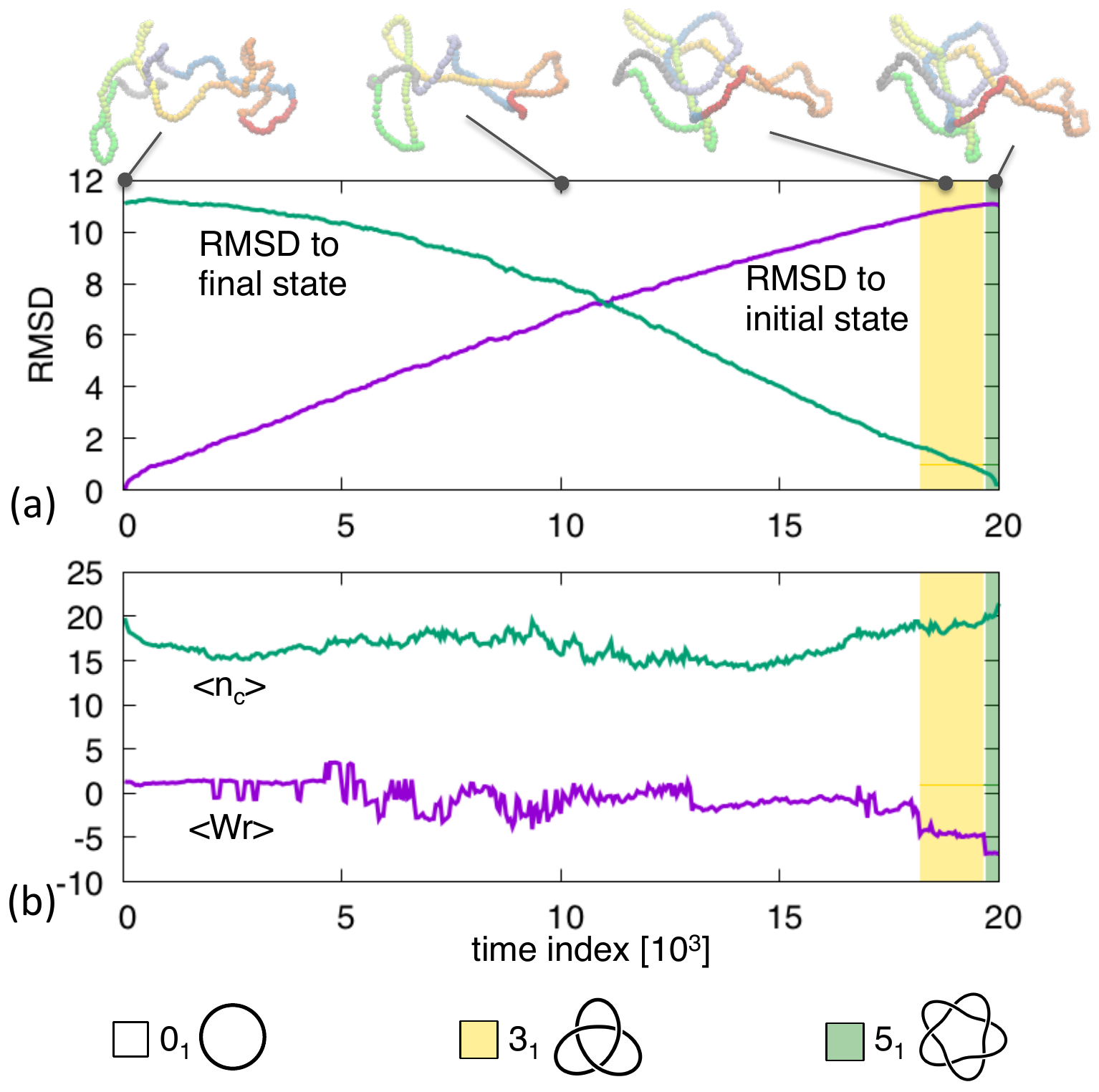}
\caption{Transition pathway between an unknotted ring and a left-handed $5_1$ knotted ring. The root-mean-square distance (RMSD) to the initial and final structures at various stages of the trajectory are shown in panel (\textbf{a}). Instantaneous configurations at selected times are highlighted. The average crossing number and writhe are shown in panel (\textbf{b}). The overlayed colored background indicates the non-trivial topological states, see legend.}
\label{fig:figure1}
\end{figure}

An overview of the typical transition pathway between these two conformations is given in Figure~\ref{fig:figure1}, where the initial unknotted and final $5_1$-knotted conformations are represented along with intermediate snapshots.

The pathway progresses steadily between these states. This is clarified by the time evolution of the root-mean-square distance (RMSD) from the start and end conformations, which progresses steadily and without lag phases, see panel (a).  Panel (b), instead, profiles other topology-related metric properties, such as the average crossing number,  $\langle n_c \rangle$, and the average writhe, $\langle Wr \rangle$.  We recall that both quantities are obtained by considering several (1000 in our case) two-dimensional projections of the oriented conformation and averaging over them a weighted sum of the projected crossings. For the crossing number each crossing carries the same $+1$ weight, while for the writhe the weight is either $+1$ or $-1$ depending on the handedness (right-hand rule) of the pair of crossings strands \cite{Adams:2004uu}.  The time evolution of the two quantities is noticeably noisier than the RMSD profile and its overall trend does not show a steady progression from initial to final state. The negative values of $\langle Wr \rangle$ in the final stages of the trajectories are consistent with the left-handedness of the target $5_1$ knot.

These properties clarify a posteriori that the imposed duration of the transition pathway is adequate: it is not so long that the conformations diffuses randomly away from the initial state before pointing towards the final state, and yet it is not so short that stochastic fluctuations are suppressed.

The associated discontinuous evolution of the topological, knotted state is highlighted by overlaid colored bands in Figure~\ref{fig:figure1}. For most of the evolution, the conformation is locked in the unknotted state and becomes non trivial only in the last $\sim$20\% of the trajectory. In this latter part, the $5_1$ state is reached via a different, intermediate topology, namely a $3_1$ knot. This is consistent with previous considerations on the unknotting number because the $3_1$ or trefoil knot has unknotting number equal to 1 and, being the simplest knot type, can optimally bridge between the $0_1$ and $5_1$ end states. {In~more general terms}, knot transitions can occur only within pairs of knots at strand passage distance equal to 1 \cite{Darcy}.

\begin{figure}[h!]
\centering
\includegraphics[width=0.6\columnwidth]{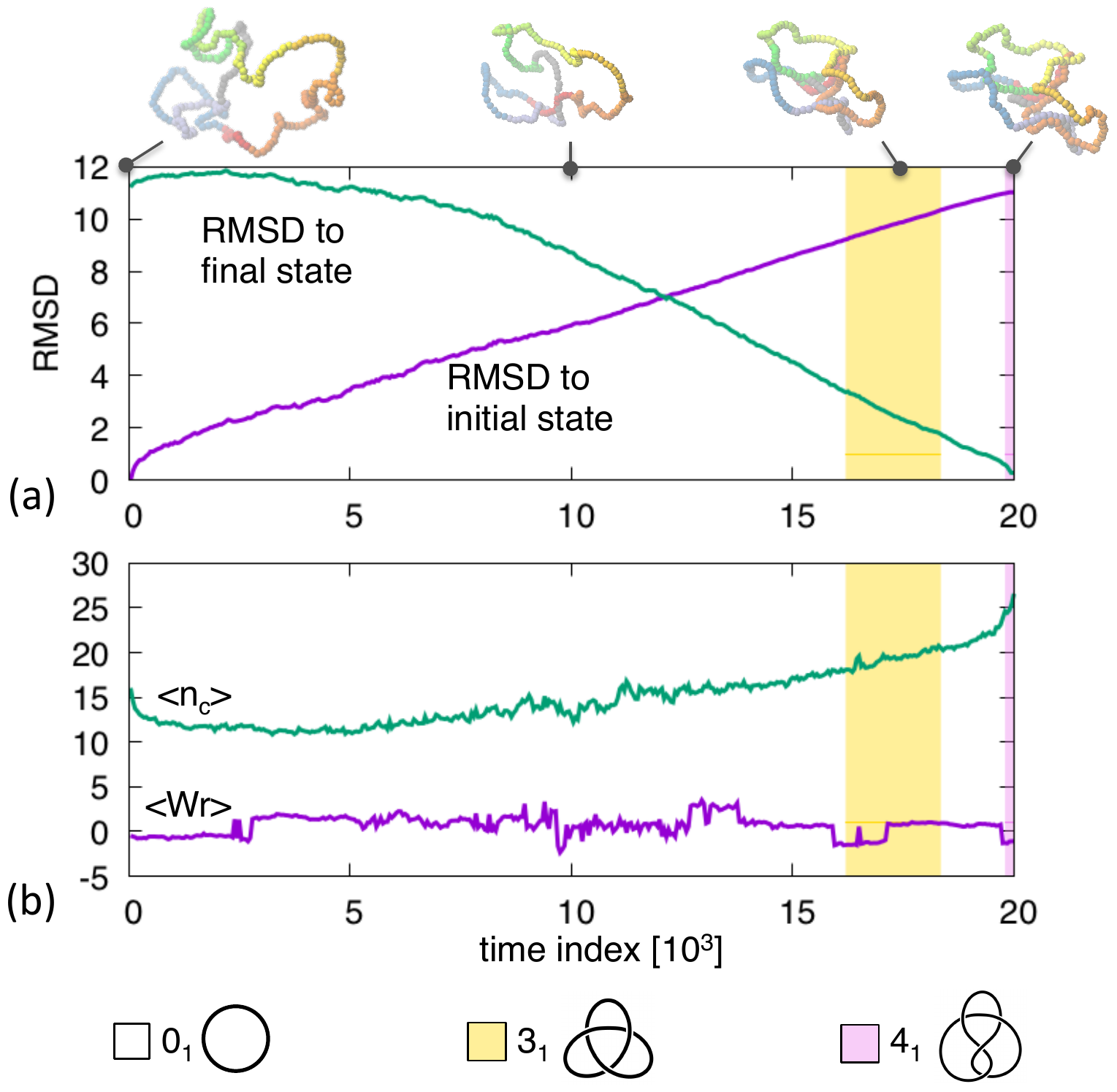}
\caption{Transition pathway between an unknotted ring and a $4_1$ knotted ring. The shown observables are the same as in Figure~\ref{fig:figure1}.}
\label{fig:figure2}
\end{figure}

This clear and intuitive progression of topological complexity is not always observed. For~instance, in Figure~\ref{fig:figure2} one notes that the pathway connecting the shown $0_1$ and $4_1$  states  switches repeatedly between unknotted and $3_1$ topologies before reaching the the target figure-of-eight one. The intermittent occupation of trefoil knots is a robust feature of $0_1 \leftrightarrow 4_1$ routes. In fact, though direct $0_1 \leftrightarrow 4_1$ are clearly possible \cite{Darcy,Flammini2004} the mediation through $3_1$ knots is observed in 10  out of 32 trajectories connecting various combinations of equilibrated initial and final states with $0_1$ and $4_1$ topologies.

The observed properties of this prototypical interconversion illustrate well the insight that can be gained from  Langevin bridging schemes and that would not be obtainable by alternative means.

For instance, allowing the system to evolve freely from the initial state would be ineffective to reach the target topology  unless it is highly represented in the canonical ensemble. {\em A fortiori} the chance that the specific target geometry is reached would practically be always negligible.

Master equation approaches based on transition rates between knot types (observed in a large number of free stochastic evolutions \cite{Flammini2004}) would be inapplicable too. Transition matrices can correctly capture that the unknot can be directly interconverted to topologies with unknotting number equal to 1 ($3_1$, $4_1$, $5_2$, $6_1$ etc.), but  the predicted Markov succession of discrete topologies, and their lifetimes, connecting  $0_1$ to $4_1$ states would have no bearings on the actual conformational evolution of ring polymers. The transition matrix approach, therefore, can elegantly recapitulate the equilibrium knotting statistics in terms of Markovian transition between topologies, giving valuable insight into the interplay of geometry and topology. However, generating viable canonical pathways connecting actual states would be beyond its scope. This is were the specificity of the proposed Langevin bridging scheme lies.

From this standpoint, particularly interesting are the transitions between equilibrated rings with different conformations but same topology. From such pathways one can understand whether  iso-topological transitions occur via pathways that maintain the same knotted state at all times. Our analysis of 270 trajectories using  the same knot type (of up to 5 crossings) for both end states, indicates that the trajectories are not constrained within a single topology.

\begin{figure}[h!]
\centering
\includegraphics[width=0.5\columnwidth]{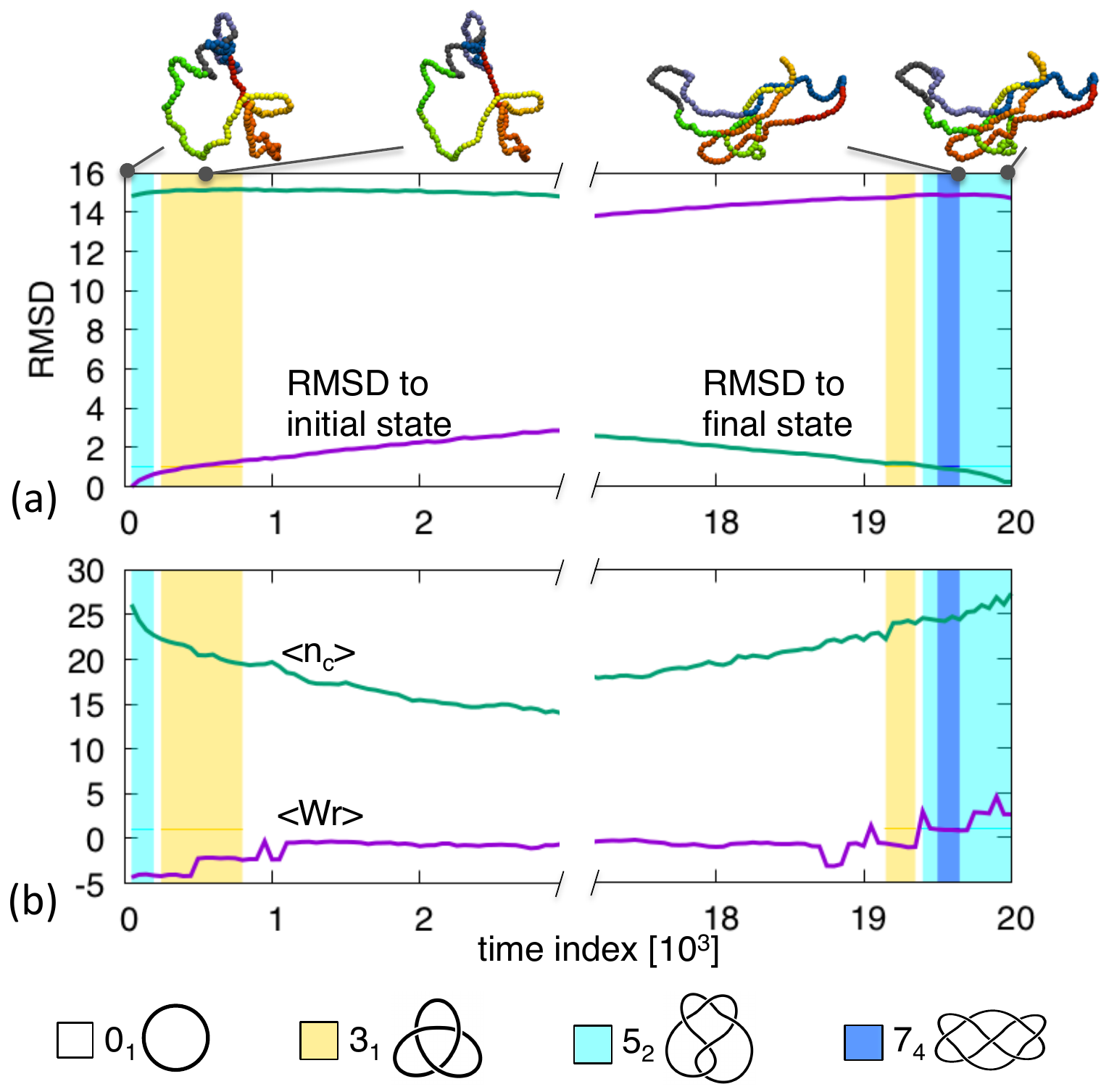}
\caption{Transition pathway between two $5_2$ knotted ring. The shown observables are the same as in~Figure~\ref{fig:figure1}.}
\label{fig:figure3}
\end{figure}

Figure~\ref{fig:figure3} shows one such trajectory with end states tied in a $5_2$ knot (same chirality). The bridging pathway clearly populates knots that are simpler ($3_1$) and more complex ($7_4$) than the initial and final topologies. The presence of $7_4$ knots on the route is particularly noteworthy because---unlike the $5_2$ one---it has unknotting number equal to 2. This means that the system evolves through states that are definitely more entangled than the initial one and these, in turn, are further simplified before the target state can be reached. This larger-than-expected intermediate complexity is frequent. In our set of 270 trajectories with end states having the same topology of up to 5 crossings, we observed that 6\% of the canonical trajectories went through states with 6 or more crossings. The most recurrent type of such knots were $6_1$, $6_2$ and the aforementioned $7_4$.

\section{Concluding Remarks}

In this study, we showed that Langevin bridging schemes provide an effective and elegant solution to the challenging problem of generating viable canonical trajectories between two assigned knotted configurations. The duration of the trajectories can also be specified a priori, thus allowing full control over the system and the simulation cost.

The method allowed us to explore transition pathways between various combination of ring conformations of all topologies up to 5 crossings. We established that such pathways often, though not always, involve intermediate topological states that are more complex than either of the connected~states.

We envisage that extensions of this scheme ought to be particularly useful to study the reconnection events that take place spontaneously in dissipative systems of fluctuating crossable filaments and flux tubes. This would require a multicanonical generalization of the approach to deal with a time-dependent number of components.

\vspace{6pt}


\section{Acknowledgments}
We thank the Beijing  Computational Science Research Center for its kind hospitality during the first stages of this work.

\section{Appendix}

\subsection{Derivation of the bridge equation}

We assume that the system is driven by a force $F(x,t)$
and is subject to stochastic dynamics in the form of an overdamped
Langevin equation.

For the sake of simplicity, we illustrate the method on a one-dimensional
system, the generalization to higher dimensions or larger number of
degrees of freedom being straightforward. We follow closely the presentation
given in Ref. \cite{orland2011generating}.

The overdamped Langevin equation reads

\begin{equation}
\frac{dx}{dt}=\frac{1}{\gamma}F(x(t),t)+ \eta(t)\label{eq:app-langevin}
\end{equation}
where $x(t)$ is the position of the particle at time $t$, driven
by the force $F(x,t)$, $\gamma$ is the friction coefficient, related
to the diffusion coefficient $D$ through the Einstein relation $D=k_{B}T/\gamma$,
where $k_{B}$ is the Boltzmann constant and $T$ the temperature
of the heat bath. In addition, $\eta(t)$ is a Gaussian white noise
with moments given by

\begin{equation}
\langle\eta(t)\rangle=0\label{eq:noise1}
\end{equation}
\begin{equation}
\langle\eta(t)\eta(t')\rangle=\frac{2 k_{B}T}{\gamma} \, \delta(t-t')\label{eq:app-noise2}
\end{equation}

The probability distribution $P(x,t)$ for the particle to be at point
$x$ at time $t$ satisfies a Fokker-Planck equation \cite{VanKampen92, Zwanzig01},

\begin{equation}
\frac{\partial P}{\partial t}=D\frac{\partial}{\partial x}\left(\frac{\partial P}{\partial x}-\beta FP\right)\label{eq:app-FP}
\end{equation}
where $\beta=1/k_{B}T$ is the inverse temperature. This equation
is to be supplemented by the initial condition $P(x,0)=\delta(x-x_{0})$,
where the particle is assumed to be at $x_{0}$ at time $t=0$. To
emphasize this initial condition, we will often use the notation $P(x,t)=P(x,t|x_{0},0)$.

We now study the probability over all paths starting at $x_{0}$ at
time $0$ and conditioned to end at a given point $x_{f}$ at time
$t_{f}$, to find the particle at point $x$ at time $t\in[0,t_{f}]$.
This probability can be written
as
\[
\mathcal{P}(x,t)=\frac{1}{P(x_{f},t_{f}|x_{0},0)}Q(x,t)P(x,t)
\]
where we use the notation
\[
P(x,t)=P(x,t|x_{0},0)
\]
\[
Q(x,t)=P(x_{f},t_{f}|x,t)
\]

Indeed, the probability for a path starting from $(x_{0},0)$ and
ending at $(x_{f},t_{f})$ to go through $x$ at time $t$ is the
product of the probability $P(x,t|x_{0},0)$ to start at $(x_{0},0)$
and to end at $(x,t)$ by the probability $P(x_{f},t_{f}|x,t)$ to
start at $(x,t)$ and to end at $(x_{f},t_{f})$.

The equation satisfied by $P$ is the Fokker-Planck equation mentioned
above (\ref{eq:app-FP}), whereas that for $Q$ is the so-called reverse
or adjoint Fokker-Planck equation  \cite{VanKampen92, Zwanzig01} given by

\begin{equation}
\frac{\partial Q}{\partial t}=-D\frac{\partial^{2}Q}{\partial x^{2}}-D\beta F\frac{\partial Q}{\partial x}\label{eq:app-FTadj}
\end{equation}

It can be easily checked that the conditional probability $\mathcal{P}(x,t)$
satisfies a new Fokker-Planck equation

\[
\frac{\partial\mathcal{P}}{\partial t}=D\frac{\partial}{\partial x}\left(\frac{\partial\mathcal{P}}{\partial x}-\left(\beta F+2\frac{\partial\ln Q}{\partial x}\right)\mathcal{P}\right)
\]

Comparing this equation with the initial Fokker-Planck (\ref{eq:app-FP})
and Langevin (\ref{eq:app-langevin}) equations, one sees that it can
be obtained from a Langevin equation with an additional potential
force

\begin{equation}
\frac{dx}{dt}=\frac{1}{\gamma}F+2D\frac{\partial\ln Q}{\partial x}+ \eta(t)\label{eq:app-bridge1}
\end{equation}

This equation has been previously obtained using the Doob transform
\cite{Doob57, Fitzsimmons92} in the probability literature and provides a simple recipe
to construct a \emph{generalized bridge}. It generates Brownian paths,
starting at $(x_{0},0)$ conditioned to end at $(x_{f},t_{f})$, with
unbiased statistics. It is the additional term $2D\frac{\partial\ln Q}{\partial x}$
in the Langevin equation that guarantees that the trajectories starting
at $(x_{0},0)$ and ending at $(x_{f},t_{f})$ are statistically unbiased.
This equation can be easily generalized to any number of degrees of freedom.

Equation (\ref{eq:app-bridge1}) is straightforwardly generalized to systems with many degrees of freedom.
For systems comprising $N$ particles interacting via a potential $U$ and subject to an
external force ${\bf F}_n$ acting on particle $n$, the evolution of the position vector ${\bf r}_n$ of the $n$th particle, is given by:
\begin{equation}
\frac{d {\bf r}_{n}}{dt}=-\frac{1}{\gamma} \grad_{{\bf r}_n} U +\frac{1}{\gamma}{\bf F}_n(t)+2D \grad_{{\bf r}_n} \ln Q +{\veceta}_{n}(t) \label{eq:app-bridge2-1}
\end{equation}
where  $Q(\left\{ {\bf r}_{n}\right\} ,t)=P\left(\{ {\bf r}_{n}^{(f)}\} ,t_{f}\mid \{ {\bf r}_{n}\},t\right)$,
$\{ {\bf r}_{n}^{(f)}\}$ is the final configuration of the system,
and the Gaussian noise ${\veceta}_{n}(t)$ satisfies
\begin{eqnarray}
&\langle\eta_{n}^{(\alpha)}(t)\rangle=0\label{eq:app-noise1-1}, \nonumber \ \ \
&\langle\eta_{n}^{(\alpha)}(t)\eta_{n'}^{(\alpha')}(t')\rangle=\frac{2 \, \kappa_b T}{\gamma}\delta_{nn'}\delta_{\alpha\alpha'}\delta(t-t')\ \nonumber
\end{eqnarray}
where $\alpha$ labels the Cartesian coordinates $x,y,z$.

{\bf Polymer chain.}
In the following we specialize eq.~(\ref{eq:app-bridge2-1}) to the case of ring polymers that freely evolve under the action of the following inter-monomer potential, $U$
\begin{equation}
\beta U=\frac{3}{2a^{2}}\sum_{n=1}^{N}\left({\bf r}_{n+1}- {\bf r}_{n}\right)^{2}+\frac{K}{2}\sum_{n=1}^{N}\left( {\bf r}_{n+1}-2 {\bf r}_{n}+ {\bf r}_{n-1}\right)^{2}\nonumber  \label{eq:app-hamiltonian}
\end{equation}
where $\beta$ is the inverse temperature and, since the chain is a ring, $ {\bf r}_{N}= {\bf r}_{0}$ and $ {\bf r}_{N+1}= {\bf r}_{1}$.

To model chains with preassigned root-mean-square bond length, $b$, and  persistence length, $l_{P}$, the bare parameters $a$ and $K$ must be set as follows:
\begin{eqnarray}
l_{P}&=&\sqrt{\frac{K a^2}{3}} \label{eq:app-bond} \\
b^{2}&=&\frac{1}{N}\langle\sum_{n=1}^{N}\left( {\bf r}_{n+1}- {\bf r}_{n}\right)^{2}\rangle\\ \nonumber
&=& \frac{a^2}{N}\sum_{p=0}^{N-1} \left[1+\frac{2Ka^2}{3}(1-\cos\omega_{p}) \right]^{-1} \nonumber \\
\end{eqnarray}
\noindent where $\omega_{p}=\frac{2\pi}{N}p$. In the limit $N\to \infty$, we have
\begin{equation}
b^2 = a^2 \int_{-\pi}^{+\pi} \frac {d\omega}{2 \pi} \frac{1}{1 + \frac{2 K a^2}{3} (1 - \cos \omega)}
\end{equation}
and for large $K$, it can be written as
\begin{equation}
b^2 =a^2\int_{-\pi}^{+\pi} \frac {d\omega}{2 \pi} \frac{1}{1 + \frac{K a^2 \omega^2}{3} } \ .
\end{equation}
After some calculations, we obtain
\begin{equation}
l_P = \frac{2}{3} K b^2
\end{equation}
which shows that the persistence length $l_P$ is proportional to the parameter $K$. In addition, the parameter $a$ is related to the Kuhn length $b$ by
\begin{equation}
a= b \sqrt {2 l_P} \ .
\end{equation}

For the considered polymer case, the Langevin bridge equation of (\ref{eq:app-bridge2-1}) is best expressed in Fourier space:
\begin{equation}
\frac{d \tilde{ {\vecrho}_p}}{dt} = -\Omega_p \tilde{{ \vecrho}}_p + \frac{D}{N} \grad_{\tilde{\vecrho}_p} \ln Q + \tilde{\veceta}_p\ ,
\end{equation}
where
\begin{eqnarray}
\tilde{\vecrho}_{p}&=&\frac{2}{N}\sum_{n=1}^{N}\cos (\omega_{p}n)\,{\bf r}_{n}  \label{eq:app-fourier_inv}\\
\Omega_p &=& (3/a^2)\, (1-\cos \omega_p) + 2K (1-\cos \omega_p)^2
\end{eqnarray}
and  $\tilde{\veceta}_p$ are the Fourier series of ${\veceta}_n(t)$ and are thus Gaussian white noises, defined by their moments
\begin{eqnarray}
\langle \tilde \eta_p(t) \rangle &=&0 \\
\langle \tilde \eta_0^{(\alpha)}(t) \eta_p^{(\alpha')}(t')\rangle &=&\frac{2D}{N} \delta_{p0}\delta_{\alpha \alpha'}\delta(t-t') \\
\langle \tilde \eta_p^{(\alpha)}(t) \eta_{p'}^{(\alpha')}(t') \rangle &=&\frac{D}{N} \delta_{pp'} \delta_{\alpha \alpha'} \delta(t-t')\ .
\end{eqnarray}

The Green's function $Q(\tilde{\vecrho}_{p},t)$ can be computed exactly
by solving the Langevin equation in Fourier space.

The calculation yields
\begin{equation}
Q(\tilde{\vecrho}_{p},t)=\exp \Bigg(-\frac{N}{D}\Omega_p\frac{ \left(\tilde {\vecrho}_p^{(f)} -\tilde {\bf R}_p(t) \right)^2}{1-e^{-2 \Omega_p (t_f-t)} }\ ,
\Bigg)
\end{equation}

so that the bridge equations then become
\begin{eqnarray}
\frac{d \tilde {\vecrho}_0}{dt} &=& \frac{1}{\gamma} \tilde {\bf F}_0(t)+\frac{\tilde {\vecrho}_0^{(f)}-\tilde {\bf R}_0(t)}{t_f-t}+\tilde {\veceta}_0(t) \nonumber  \\
\frac{d \tilde {\vecrho}_p}{dt} &=&\frac{1}{\gamma} \tilde {\bf F}_p(t) -\Omega_p \tilde {\vecrho}_p(t)+ {\Omega_p}
\frac { \tilde {\vecrho}_p^{(f)}- \tilde {\bf R}_p(t) }{\sinh \Omega_p (t_f-t)}
+
\tilde \veceta_p(t) \nonumber
\end{eqnarray}
\noindent  where  $\tilde {\vecrho}_p^{(f)}$ denotes the final configuration of the chain in Fourier components,
\begin{eqnarray}
\tilde {\bf R}_0(t) &=& \tilde {\vecrho}_0(t) + \frac{1}{\gamma} \int_t ^{t_f} d\tau \tilde {\bf F}_0(\tau) \nonumber \\
\tilde {\bf R}_p(t) &=& \tilde {\vecrho}_p (t) e^{-\Omega_p (t_f-t)} + \frac{1}{\gamma} \int_t^{t_f} d\tau e^{-\Omega_p (t_f- \tau)} \tilde {\bf F}_p(\tau) \nonumber
\end{eqnarray}
and the forces $\tilde {\bf F}_p(t)$ are the Fourier series of the forces ${\bf F}_n(t)$ defined according to eq. (\ref{eq:app-fourier_inv}). Note that these equations bear some resemblance to the bridge equations for an Ornstein-Uhlenbeck process \cite{majumdar2015effective}, since the original Langevin equations are linear in both cases.

These equations can be discretized and solved numerically, from an initial configuration $\tilde \vecrho_p^{(0)}$ to a final one $\tilde \vecrho_p^{(f)}$.




\bibliographystyle{mdpi}

\renewcommand\bibname{References}

\end{document}